# Irreversible Effects in LaGa$_{1-x}$Mn$_x$O$_3$

N. Noginova, F. Chen, E. Etheridge.
*Norfolk State University, Norfolk, VA*

*Abstract.* Quasi-irreversible increase in the electrical conductivity is observed in single crystals of LaGa$_{1-x}$Mn$_x$O$_3$. The effect lasts for long time at room temperature and can be erased by heating of the crystal above the phase transition temperature. We explain the observed effects in terms of ionization and local lattice distortion processes.

## INTRODUCTION

In the past few years, perovskite manganites (such as La$_{1-x}$Sr$_x$MnO$_3$, La$_{1-x}$Ca$_x$MnO$_3$, *etc.*) have attracted much scientific attention due to effect of the colossal magnetoresistance (CMR) effect and complex diagram of phase transitions in these materials [1]. As was recently reported, some low-doped manganites exhibit effect of irreversibility in transport and magnetic properties induced by optical illumination, thermal, electric or magnetic field cycling in the range of ferromagnetic transition [2-7]. These memory effects are explained with intrinsic coexistence of two or more phases with different magnetic and transport properties on the submicrometer scale and microstructural changes in the phases content.

On the other hand, high-resistivity perovskite materials with very low concentration of Mn ions also exhibit interesting memory effects. Mn doped yttrium orthoaluminates, Mn:YAlO$_3$, high-quality optical single crystals, demonstrate significant changes in color and photorefractive index induced by photo illumination [8, 9]. These effects are quasi-permanent, lasting for many years at room temperature, and completely erasable by heating the crystal up to ~ 230 C.

To better understand the origin of the irreversibility in perovskites with different content of Mn ions, we have studied the series of LaGa$_{1-x}$Mn$_x$O$_3$ with concentration of Mn ions varying from 2% to 100%. In contrast to the well-known manganite CMR systems, in our single crystal materials Mn ions are diluted by nonmagnetic Ga ions, which enter the same place in the crystal structure as Mn ions. As was reported earlier [10], the low-field conductivity of such systems can be described in terms of the small polaron hopping model. Here we mostly concentrate on non-linearity and irreversible effects observed in the electric conductivity.

## EXPERIMENT

Crystals of LaGa$_{1-x}$Mn$_x$O$_3$ with x≤ 0.2 were grown by the Czochralski technique in a slightly oxidizing atmosphere. Systems LaGa$_{1-x}$Mn$_x$O$_3$ with higher concentration 0.5 ≤ x ≤ 1 were fabricated by floating zone technique. According to the X-ray diffraction analysis, LaMnO$_3$ has a distorted orthorhombic *Pbnm* crystal lattice of O' type ($c/\sqrt{2}$ < a, b). With decrease in Mn concentration, the lattice changes from the O' type at 100% of

Mn to O* type (a, b~$c/\sqrt{2}$) at 50% Mn or less. These changes are related to the gradual transition from the static JT effect in materials with high concentration of Mn ions to the dynamic JT in the systems with intermediate and low concentration of Mn. [11,12]. With increase in temperature, materials undergo phase transformation to rhombohedral or quasi-cubic phase. The temperature and type of the phase transformation depends on the concentration of Mn ions, varying from ~400K at materials with x= 0.005 to ~750K at x=1.

The LaMnO$_3$ crystal is antiferromagnetic with T$_c$ ≈ 130K. Diluted materials with x = 0.8, 0.5 demonstrate ferromagnetic transition, with T$_c$ ≈125 K and 90 K correspondingly. The ferromagnetic transition was not observed in the range of temperatures studied (4-300K) in the crystals that had lower concentration of Mn. Mn ions are expected to occupy the Ga$^{3+}$ sites and be in 3+ valence state. However, the spectroscopic studies in 0.5% and 2% doped samples indicated also a small amount of different valence states, such as Mn$^{2+}$, Mn$^{4+}$, and Mn$^{5+}$ [13].

Measurements of the electrical conductivity were performed in the two contact scheme using crystal plates with the thickness of 0.3 - 5 mm for samples with x = 0.05≤ x ≤0.8. At room temperature, the materials in study are highly insulating. The resistance of the samples is in the range of tens or hundreds of MΩ, strongly exceeding possible contact resistances. Four point contact technique has been used in the experiments with x≥0.8.

Electron Spin Resonance (ESR) spectra have been recorded to characterize magnetic behavior of the samples. Bruker AMX spectrometer has been used in the experiments.

**RESULTS**

In the quasi-linear regime, at low enough currents (j < 10$^{-3}$ A/cm$^2$ for x≥0.5, and j < 10$^{-5}$ A/cm$^2$ for x ≥0.05), the electrical conductivity of all the systems in study can be described adequately with non-adiabatic approximation of the hopping model with the values of the activation energy and resistivity coefficient strongly dependent on the concentration of Mn ions [10],

$$\rho = \rho_0 T^{3/2} \exp(E_g / k_B T). \tag{1}$$

where $\rho_0$ is the resistivity coefficient, $T$ is the temperature, $k_B$ is the Bolzmann constant, $E_g$ is the activation energy ranging from 80 meV for x=1 to 600 meV for x= 0.005.

To study non-linear and hysteresis effects at higher currents, we measured the voltage-current dependences for currents of 0.001-100 mA in the constant current regime for the concentrations 0.1≤ x ≤1. In the experiments, the IV curves were recorded during a round-trip scanning in the range of (0, I$_m$) and subsequent (0, -I$_m$).

A typical IV curve is shown in Fig. 1. As one can see, the voltage initially grows with the increase of the current, and reaches maximum at some particular J$_m$. The further increase of the current leads to the decrease of the voltage (switching to the negative differential resistance, $R_d = dV/dI < 0$). During decrease of the current, the voltage is lower then during the increase of the current. Note that the hysteresis loop at the first half-scan is wider that that during the second.

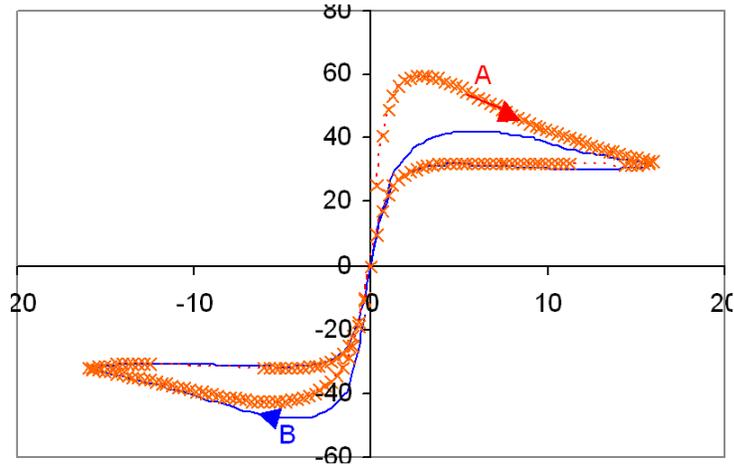

Fig. 1. Non-linearity and hysteresis in voltage-current curves in $LaGa_{0.2}Mn_{0.8}O_3$. Arrows show initial directions of the scans. Scan B was taken 10 minutes later after the Scan A.

Such behavior can be explained partially by non-equilibrium heating and strong temperature dependence of the resistance. However, at a very slow rate (4 hours run), when the sample is in thermal equilibrium, the loop becomes narrow but still exists.

Non-linearity and hysteresis effect were observed for all Mn concentrations in study, including $LaMnO_3$. Voltage-current curves for different concentrations recorded at the first run from 0 to $I_m$ are shown in Fig. 2. Switching to the lower resistance was observed at some particular $J_o$ depending on the Mn concentration. It accompanied by the irreversible effects in conductivity: after applications of high currents the low field conductivity becomes much higher than that before.

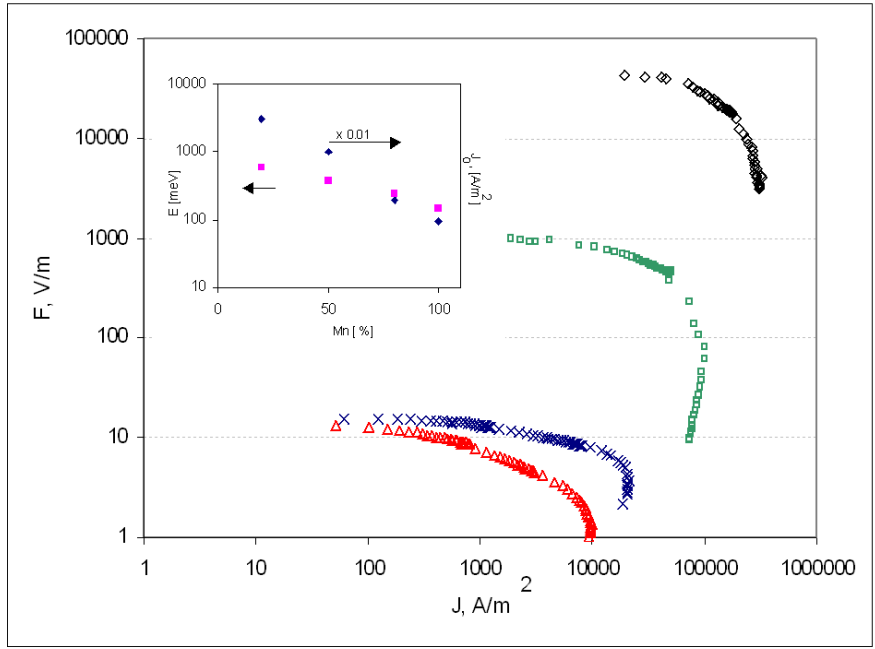

Fig. 2. Voltage-current dependences in $LaGa_{1-x}Mn_xO_3$ at x = 1 (triangles), x=0.8 (crosses), x=0.5 (squares), and x=0.2 (diamonds). Insert: Parameters of the charge transport *vs* Mn concentration. $E_g$ (squares), $J_o$ (diamonds)

Fig. 3 demonstrates the voltage-current curves in low current range taken in $LaGa_{0.2}Mn_{0.8}O_3$ before and after application of 50 mA dc current. The voltage *vs* current dependence measured after the application of the high current has much lower slope comparing to that measured initially. This corresponds to the substantial increase in conductivity twofold or threefold, depending on a particular sample. The resistance, R, increases rapidly during first several minutes to a certain level but still is much lower than the resistance, $R_0$, before the application of high currents.

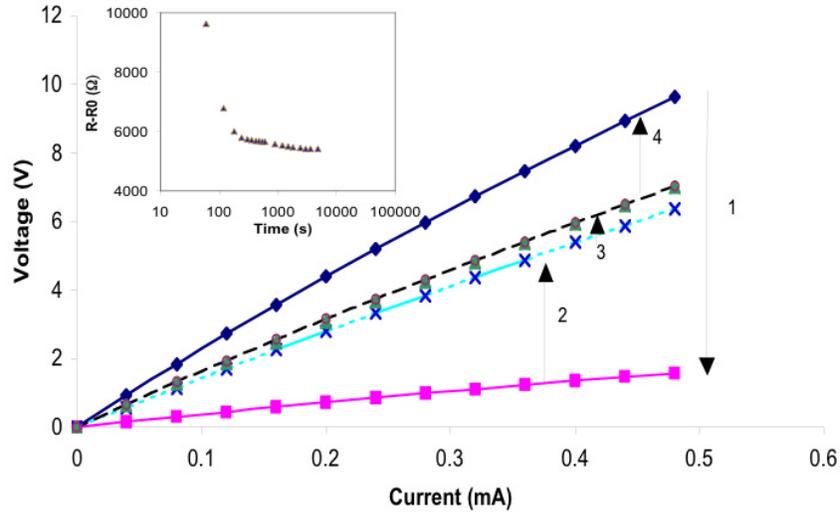

Fig. 3. Voltage vs current dependences in LaGa$_{0.2}$Mn$_{0.8}$O$_3$ (low currents range), measured before (diamonds) application of 50 mA, right after (cubes), two minutes later (crosses), thirty minutes later (triangles), and 2 hours later (circles). Arrows show directions of the curve evolution as an effect of high-current application (1), fast recovering (2), slow relaxation (3) and high temperature treatment (4). Inset: Recovering of the resistance after application of a high current.

The recovering of the low field resistance in the dependence on time is shown in inset of Fig. 3. The initial fast increase of the resistance may be attributed to the cooling of the crystal heated by the high currents. The relaxation process in the long time range is much more slow and follows approximately logarithmic time dependence.

Very similar to the situation in Mn doped aluminates, the induced changes last for a long time at room temperature and can be easily erased by heating. After heating the crystal at 260 C during 20 minutes, and subsequent cooling down, the measured conductivity was practically the same as that before the application of high currents.

In contrast to the electrically induced irreversibility observed for all concentrations in study and in the broad temperature range, irreversibility in magnetic behavior is observed in the narrow range of the ferromagnetic transition, only for one concentration of x=0.8.

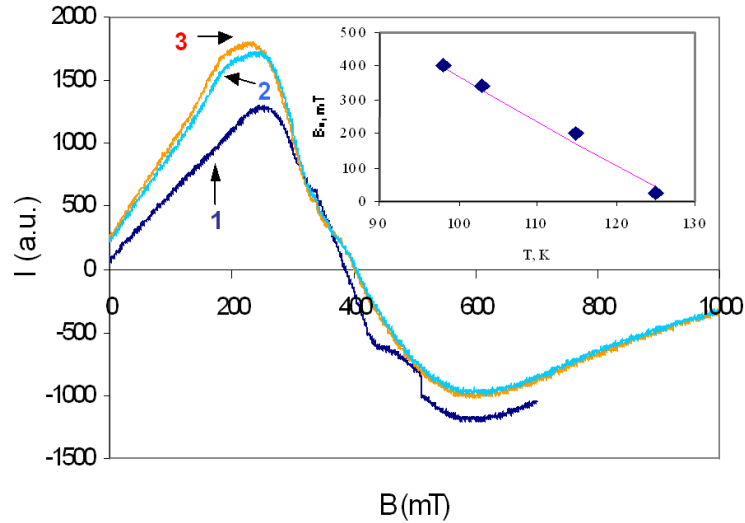

Fig. 4. ESR signal of field cooled LaGa$_{0.2}$Mn$_{0.8}$O$_3$ sample (H$_{FC}$ = 0.7 T) at the first (Trace 1) and subsequent runs (Traces 2, 3). T=103 K. Inset: Temperature dependence of the switching field.

As shown in Fig. 4, the ESR line in zero-field cooled LaMn$_{0.8}$Ga$_{0.2}$O$_3$ recorded at first run demonstrates abrupt jumps (Trace 1) at a particular field H$_c$, while at following runs the signal repeats itself (Traces 2, 3)). The field where first jump is observed, depends on the temperature (see inset on Fig. 4) as H$_c$ ~ 1-T/T$_c$ with T$_c$ =128 K.

**DISCUSSION**
Many manganites, in particular, systems with low content of Mn$^{4+}$ ions, can be considered as a submicron scale mixture of two or more phases, including ferromagnetic conductive nanodroplets of phase with Mn$^{3+}$ and Mn$^{4+}$ ions, and highly insulating antiferromagnetic phases with Mn$^{3+}$. In our materials LaGa$_{1-x}$Mn$_x$O$_3$, content of the insulating phases is much higher than in CMR materials due to the absence of specially introduced divalent doping and a partial substitution of the Mn with Ga. Low field conductivity in such systems has percolation character and associated with the presence of small amount of holes due to defects.

As it was discussed in [10], among possible sources of non-linearity in dilute manganites are heating and an effect of the electric field on the hopping probability, which arises when the additional energy of a carrier due to the electric field, *F*, becomes comparable to $k_BT$. Besides, it is reasonable to assume that the distribution of the current density in the sample can be highly non-uniform, and changes with increase of the temperature and electric field. Taking into account these effects qualitatively well explains the shape of the IV curves. However, obviously, this model cannot explain the irreversible changes in conductivity observed in experiment.

To explain the irreversible effects in the low-field conductivity induced by high currents, let us point to strong similarity between the memory effects in electric conductivity in LaGa$_{1-x}$Mn$_x$O$_3$ and photoinduced coloration and refractive index change in Mn:YalO$_3$. In both cases the irreversible effects are apparently related to the ionization of Mn ions and redistribution of charge. In both cases, the memory effects last for a long time at room temperature and demonstrate very slow logarithmic relaxation in a similar time scale

(half-lifetime is ~ 122 years in Mn:YalO$_3$ [8], and ~300 years in LaGaMnO$_3$). In similarity, both effects are completely erasable by heating the crystals to a temperature of 230-300 C, which is higher than the phase transformation temperature.

We believe that in both cases, the effects of lattice distortion play the major roles in photo or electrically induced irreversibility. In particular, in LaGa$_{1-x}$Mn$_x$O$_3$, high currents are sources of both thermo- and field induced ionization of Mn ions from Mn$^{3+}$ to Mn$^{4+}$ or Mn$^{5+}$ valence states. The oxidation of Mn$^{3+}$ results in the local transformation of the corresponding lattice cells and removal of their JT distortions. This process is sped up if the temperature is close to the phase transformation.

Just after application of high currents, recombination of excess carriers leads to the resistance growth while the crystal is cooling down. At room temperature, the recombination process is very slow due to an additional energy needed for creation of JT Mn$^{3+}$ ions. Heating of the sample to the phase transformation temperature leads to the recombination of the Mn$^{4+}$ ions and electrons (production of Mn$^{3+}$ ions), and corresponding decrease of the carrier concentration in the material.

In contrast to the electrically induced irreversibility observed in broad range of Mn concentrations including relatively low concentrations, magnetic effects are observed only for high concentrations of Mn ions, when materials can be considered to be a mixture of ferromagnetic and antiferromagnetic phase (see also [3]). The magnetization jumps and the effect of "magnetic training" are typical for such mixtures [3, 14] and can be explained with reorientation and collapsing of antiferromagnetic domains.

In conclusion, irreversibility in electrical conductivity induced by applications of high currents was studied in single crystals of LaGa$_{1-x}$Mn$_x$O$_3$. Similar to the optically induced irreversibility in low doped perovskites, these effects last for a long time at room temperature and can be erased by the heating of the crystal above the phase transition temperature. This quasi-irreversible increase in conductivity can be related to ionization of Mn ions and subsequent lattice distortion processes.

**AKCNOWLEDGMENTS**

Research was supported by NSF CREST Project HRD-9805059.